\documentclass[aps,prl,preprint,tightenlines,superscriptaddress,showpacs]{revtex4}

\usepackage{graphicx}
\usepackage{dcolumn}
\usepackage{bm}

\newcommand{\Br}{\mathcal{B}}
\newcommand{\bsgamma}{b \to s\gamma}
\newcommand{\coshel}{\cos \theta_{\mathrm{hel}}}
\newcommand{\DE}{\Delta E}
\newcommand{\Ebeam}{E_{\mathrm{beam}}^*}
\newcommand{\GeV}{\mathrm{~GeV}}
\newcommand{\Kpi}{K^+\pi^-}
\newcommand{\Kpipi}{K^+\pi^-\pi^+}

\newcommand{\Ktwost}{K_2^*(1430)}
\newcommand{\Mbc}{M_{\mathrm{bc}}}
\newcommand{\MeV}{\mathrm{~MeV}}
\newcommand{\MKX}{M_{K_X}}

\newcommand{\fbi}{\mathrm{fb}^{-1}}
\newcommand{\qq}{q\bar{q}}

\newcommand{\NBBmillunit}{31.9}
\newcommand{\LumONRES}{29.4}

\newcommand{\YkpigammaOneFourN}{27 {\,}^{+8}_{-7} {\,}^{+1}_{-3}}
\newcommand{\YkpigammaOneFourNstr}{%
   27 {\,}^{+8}_{-7} \mbox{(stat.)} {\,}^{+1}_{-3} \mbox{(syst.)}}
\newcommand{\SNFkpigammaOneFourN}{5.0} 

\newcommand{\SNFktwostgammaN}{3.2} 
\newcommand{\YktwostgammaN}{21 {\,}^{+8}_{-7} {\,}^{+0}_{-1}}
\newcommand{\EFFktwostgammaN}{(5.0 \pm 0.3)\%}
\newcommand{\EFFktwostgammaNtbl}{5.0 \pm 0.3}
\newcommand{\BFktwostgammaNtbl}{1.3 \pm 0.5 \pm 0.1}
\newcommand{\BFktwostgammaNstr}{%
  ( 1.3 \pm 0.5 \mbox{(stat.)} \pm 0.1 \mbox{(syst.)} ) \times 10^{-5}}
\newcommand{\YkstAgammaNnosyst}{7.7 {\,}^{+7.1}_{-5.7} {\,}^{+0.5}_{-1.3}}
\newcommand{\YULkstAgammaN}{19}
\newcommand{\EFFkstAgammaNtbl}{0.58 \pm 0.12} 
\newcommand{\BFULkstAgammaNtbl}{13} 
\newcommand{\YkpinresNnosyst}{0.0 {\,}^{+4.6}_{-0.0} \pm 0.0}
\newcommand{\YULkpinresN}{15}
\newcommand{\EFFkpinresNtbl}{19 \pm 1}
\newcommand{\BFULkpinresNtbl}{0.26}

\newcommand{\EFFkpitotalNtbl}{18 \pm 2}
\newcommand{\BFkpitotalNtbl}{0.46 {\,}^{+0.13}_{-0.12} {\,}^{+0.05}_{-0.07}}
\newcommand{\SNFkpipigamma}{5.9} %

\newcommand{\Ykpipigamma}{57 {\,}^{+12}_{-11} {\,}^{+6}_{-2}}
\newcommand{\Ykpipigammastr}{%
  57 {\,}^{+12}_{-11} \mbox{(stat.)} {\,}^{+6}_{-2} \mbox{(syst.)}}
\newcommand{\kstpiMCFRACkoneB}{0.74 \pm 0.14}
\newcommand{\krhoMCFRACkoneA}{0.68 \pm 0.17}
\newcommand{\SNFkstpigamma}{3.7} 
\newcommand{\SNFkrhogamma}{2.2} 
\newcommand{\SNFkstpiANDkrhogamma}{6.2}
\newcommand{\Ykstpigamma}{33 {\,}^{+11}_{-10} \pm 2}
\newcommand{\Ykrhogamma}{24 \pm 12 {\,}^{+4}_{-7}}
\newcommand{\Ynresnosyst}{0 {\,}^{+11}_{-0} \pm 0}
\newcommand{\YULkrhogamma}{43}
\newcommand{\YULnres}{20}

\newcommand{\EFFkpipigammatbl}{7.5 \pm 0.7}
\newcommand{\EFFkstpigammatbl}{5.0 \pm 0.5}
\newcommand{\EFFkrhogammatbl}{7.4 \pm 0.7}
\newcommand{\EFFnrestbl}{7.6 \pm 0.7}
\newcommand{\BFkpipigammastr}{%
 ( 2.4 \pm 0.5 \mbox{(stat.)}
   {\,}^{+0.4}_{-0.2} \mbox{(syst.)}) \times 10^{-5}}

\newcommand{\BFkpipigammatbl}{2.4 \pm 0.5 {\,}^{+0.4}_{-0.2}}
\newcommand{\BFkstpigammatbl}{2.0 {\,}^{+0.7}_{-0.6} \pm 0.2}
\newcommand{\BFkrhogammatbl}{1.0 \pm 0.5 {\,}^{+0.2}_{-0.3}}
\newcommand{\BFULkrhogammatbl}{2.0}
\newcommand{\BFULnrestbl}{0.92}
\newcommand{\NSBOXkoneAgamma}{6}
\newcommand{\NBGkoneAgamma}{2.0 \pm 0.6}   
\newcommand{\YkoneAgamma}{4.0 \pm 2.4 \pm 0.6} 
\newcommand{\YULkoneAgamma}{10}
\newcommand{\EFFkoneAgammatbl}{0.40 \pm 0.08} 
\newcommand{\BFULkoneAgammatbl}{9.9} 
\newcommand{\YkxOneFourgamma}{26 \pm 6 {\,}^{+2}_{-0}}
\newcommand{\YULkxOneFourgamma}{36}
\newcommand{\EFFXkoneBgammatbl}{2.6 \pm 0.3} 
\newcommand{\BFULXkoneBgammatbl}{5.0}  

\begin{document}

\vspace*{-3\baselineskip}
\resizebox{!}{3cm}{\includegraphics{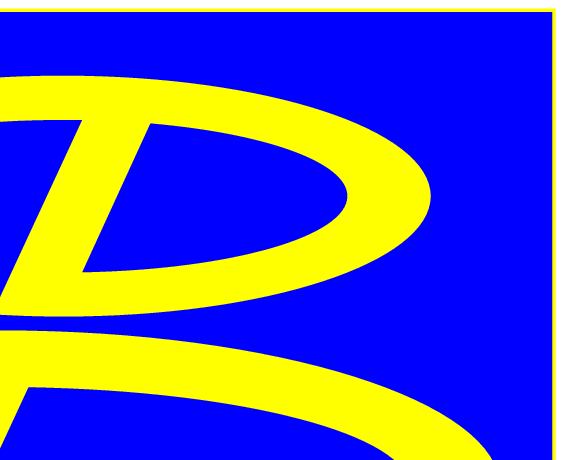}}

\preprint{BELLE Preprint 2002-10}
\preprint{KEK Preprint 2002-15}
\preprint{KUNS-1785}
\title{Radiative $B$ Meson Decays into $K\pi\gamma$
and $K\pi\pi\gamma$ Final States}





\affiliation{Aomori University, Aomori}
\affiliation{Budker Institute of Nuclear Physics, Novosibirsk}
\affiliation{Chiba University, Chiba}
\affiliation{Chuo University, Tokyo}
\affiliation{University of Cincinnati, Cincinnati OH}
\affiliation{Deutsches Elektronen--Synchrotron, Hamburg}
\affiliation{University of Frankfurt, Frankfurt}
\affiliation{Gyeongsang National University, Chinju}
\affiliation{University of Hawaii, Honolulu HI}
\affiliation{High Energy Accelerator Research Organization (KEK), Tsukuba}
\affiliation{Hiroshima Institute of Technology, Hiroshima}
\affiliation{Institute of High Energy Physics, Chinese Academy of Sciences, Beijing}
\affiliation{Institute of High Energy Physics, Vienna}
\affiliation{Institute for Theoretical and Experimental Physics, Moscow}
\affiliation{J. Stefan Institute, Ljubljana}
\affiliation{Kanagawa University, Yokohama}
\affiliation{Korea University, Seoul}
\affiliation{Kyoto University, Kyoto}
\affiliation{Kyungpook National University, Taegu}
\affiliation{Institut de Physique des Hautes \'Energies, Universit\'e de Lausanne, Lausanne}
\affiliation{University of Ljubljana, Ljubljana}
\affiliation{University of Maribor, Maribor}
\affiliation{University of Melbourne, Victoria}
\affiliation{Nagoya University, Nagoya}
\affiliation{Nara Women's University, Nara}
\affiliation{National Kaohsiung Normal University, Kaohsiung}
\affiliation{National Lien-Ho Institute of Technology, Miao Li}
\affiliation{National Taiwan University, Taipei}
\affiliation{H. Niewodniczanski Institute of Nuclear Physics, Krakow}
\affiliation{Nihon Dental College, Niigata}
\affiliation{Niigata University, Niigata}
\affiliation{Osaka City University, Osaka}
\affiliation{Osaka University, Osaka}
\affiliation{Panjab University, Chandigarh}
\affiliation{Peking University, Beijing}
\affiliation{RIKEN BNL Research Center, Brookhaven NY}
\affiliation{University of Science and Technology of China, Hefei}
\affiliation{Seoul National University, Seoul}
\affiliation{Sungkyunkwan University, Suwon}
\affiliation{University of Sydney, Sydney NSW}
\affiliation{Tata Institute of Fundamental Research, Bombay}
\affiliation{Toho University, Funabashi}
\affiliation{Tohoku Gakuin University, Tagajo}
\affiliation{Tohoku University, Sendai}
\affiliation{University of Tokyo, Tokyo}
\affiliation{Tokyo Institute of Technology, Tokyo}
\affiliation{Tokyo Metropolitan University, Tokyo}
\affiliation{Tokyo University of Agriculture and Technology, Tokyo}
\affiliation{University of Tsukuba, Tsukuba}
\affiliation{Utkal University, Bhubaneswer}
\affiliation{Virginia Polytechnic Institute and State University, Blacksburg VA}
\affiliation{Yokkaichi University, Yokkaichi}
\affiliation{Yonsei University, Seoul}

\author{S.~Nishida}           
\affiliation{Kyoto University, Kyoto}
\author{M.~Nakao}             
\affiliation{High Energy Accelerator Research Organization (KEK), Tsukuba}
\author{K.~Abe}               
\affiliation{High Energy Accelerator Research Organization (KEK), Tsukuba}
\author{K.~Abe}               
\affiliation{Tohoku Gakuin University, Tagajo}
\author{T.~Abe}               
\affiliation{Tohoku University, Sendai}
\author{Byoung~Sup~Ahn}       
\affiliation{Korea University, Seoul}
\author{H.~Aihara}            
\affiliation{University of Tokyo, Tokyo}
\author{M.~Akatsu}            
\affiliation{Nagoya University, Nagoya}
\author{Y.~Asano}             
\affiliation{University of Tsukuba, Tsukuba}
\author{T.~Aushev}            
\affiliation{Institute for Theoretical and Experimental Physics, Moscow}
% ...

\collaboration{The Belle Collaboration}
\noaffiliation


\begin{abstract}
We report observations of
radiative $B$ meson decays into the $K^+\pi^-\gamma$
and $K^+\pi^-\pi^+\gamma$ final states.
In the $B^0 \to K^+\pi^-\gamma$ channel,
we present evidence for decays via an intermediate tensor
meson state with a
branching fraction of $\Br(B^0 \to \Ktwost^0\gamma) = \BFktwostgammaNstr$.
We measure the branching fraction
$\Br(B^+ \to \Kpipi\gamma) = \BFkpipigammastr$,
in which the $B^+\to K^{*0}\pi^+\gamma$ and
$B^+ \to K^+\rho^0\gamma$ channels dominate.
The analysis is based on a
dataset of $\LumONRES~\fbi$ recorded by the Belle experiment at
the KEKB collider.
\end{abstract}



\maketitle

Since the first measurement of the inclusive branching fraction
for $B \to X_s\gamma$ by the CLEO collaboration
in 1995~\cite{CLEO:1995:bsgamma}, the flavor
changing neutral current process $b\to s\gamma$ has been used as a
sensitive probe to search for physics beyond the Standard
Model (SM).
In experiments at the $\Upsilon(4S)$, a pseudo-reconstruction
technique, in which the $X_s$ state is reconstructed from one kaon and
multiple pions, has been the most powerful tool to identify
$b\to s\gamma$ events. In order to measure more precisely the inclusive
rate, a detailed knowledge of the exclusive final states is required. In
addition to the already established
$B \to K^*\gamma$ decay~\cite{footnote:K*(892)},
there are several known resonances that can contribute to the final
state.
CLEO has reported evidence for $B \to \Ktwost\gamma$~\cite{CLEO:2000:radb}.
Some theoretical predictions for the
branching fractions of the exclusive decays
can be found in Ref.~\cite{kxgamma-prediction}.
Exclusive decays, such as $B \to K_1(1400)\gamma$,
can also be used to measure the photon helicity,
which may differ from the SM prediction in some new physics models
~\cite{Gronau:2001ng}.

In this Letter, we report on a search for resonant structures $K_X$
above the $K^*$ mass
in radiative $B$ meson decays.
The analysis is based on a
data sample of $\LumONRES~\fbi$ ($\NBBmillunit$ million $B\bar{B}$ events)
recorded by
the Belle detector~\cite{Belle:NIM} at KEKB~\cite{KEKB:unpublished}.
KEKB is an asymmetric energy $e^+e^-$
collider (3.5 GeV on 8 GeV) operated at the $\Upsilon(4S)$ resonance.
The Belle detector
has a three-layer silicon vertex detector
(SVD), 50-layer central drift chamber (CDC), an array of aerogel
Cherenkov counters (ACC), time-of-flight scintillation counters (TOF),
an electromagnetic calorimeter of CsI(Tl) crystals (ECL).

We select events that contain a high energy photon ($\gamma$)
with an energy between 1.8 and 3.4 GeV
in the $\Upsilon(4S)$ center-of-mass (CM) frame
and within the acceptance of the barrel ECL
($33^\circ<\theta_\gamma<128^\circ$).
In order to reduce the background from $\pi^0, \eta \to	\gamma\gamma$ decays,
we combine 
the photon candidate
with all other photon clusters in the event
and reject the candidate if the invariant mass of any pair is
within $18 \MeV/c^2$ ($32 \MeV/c^2$)
of the nominal $\pi^0$ ($\eta$) mass
(this condition is referred to as the $\pi^0/\eta$ veto).

We search for $K_X$ resonances decaying into
two-body ($\Kpi$) and three-body ($\Kpipi$) final
states~\cite{footnote:charge-conjugation}
in the invariant mass ($\MKX$) range up to $2.4 \GeV/c^2$.
For the $\Kpi$ final state, the range $\MKX < 1.2 \GeV/c^2$
is excluded to remove $K^*$ contributions.
Charged tracks
are required to have CM momenta greater than
$200 \MeV/c$, and
to have impact parameters
within $\pm 5 \mathrm{~cm}$
of the interaction point along the positron beam axis
and within $0.5 \mathrm{~cm}$ in the transverse plane.
To identify kaon and pion candidates, we use a likelihood
ratio that is calculated by combining information from
the ACC, TOF, and $dE/dx$ (CDC) systems.
We apply
a tight selection 
with an efficiency
(pion misidentification rate) of 83\% (8\%)
for charged kaon candidates and
a loose selection 
with an efficiency
(kaon misidentification rate) of 97\% (28\%)
for charged pion candidates.

We reconstruct $B$ meson candidates from a photon
and a $K_X$ system by
forming two independent kinematic variables: the beam constrained mass
$\Mbc \equiv \sqrt{\left(\Ebeam/c^2\right)^2
 - \left(|\vec{p}_{K_X}^{\,*}+\vec{p}_\gamma^{\,*}|/c\right)^2}$
and
$\Delta{E}\equiv E_{K_X}^* + E_\gamma^* - \Ebeam$,
where $\Ebeam$ is the beam energy,
and $\vec{p}_\gamma^{\,*}$, $E_\gamma^*$,
$\vec{p}_{K_X}^{\,*}$, $E_{K_X}^*$ are
the momenta and energies of the photon
and the $K_X$ system, respectively, calculated in the CM frame.
In order to improve the $\Mbc$ resolution, the photon
momentum is rescaled
so that $|\vec{p}_\gamma^{\,*}|=(\Ebeam-E_{K_X}^*)/c$
is satisfied.

The largest source of background originates from
continuum $\qq$ ($q = u,d,s,c$) production.
To suppress this background,
we use a Fisher discriminant~\cite{Fisher:1936et}
formed from six modified Fox-Wolfram moments~\cite{Fox:1978vu}
and the cosine of the $B$ meson flight direction ($\cos\theta_B^*$).
The moments are calculated in the rest frame of the $B$ candidate
to avoid a correlation with $\Mbc$~\cite{Belle:2001:bsgamma}.
Signal and background events are classified according to a
likelihood ratio
$LR = \mathcal{L}_\mathrm{sig}/
(\mathcal{L}_\mathrm{sig} + \mathcal{L}_\mathrm{bg})$,
where the likelihood $\mathcal{L}_\mathrm{sig}$
($\mathcal{L}_\mathrm{bg}$) is the product of
the probability density functions (PDF)
of the Fisher discriminant and $\cos\theta_B^*$
for signal (background).
The PDFs for the Fisher discriminant are determined from
Monte Carlo (MC) simulations.
For $\cos\theta_B^*$, we assume a
$1 - \cos^2\theta_B^*$ behaviour for signal events and
a flat distribution for continuum background.
The selection criteria on the likelihood ratio are chosen so that
$S/\sqrt{S+N}$ is maximized,
where $S$ and $N$ are
(MC) signal and background yields, respectively.
The optimized criteria retain 68\% of the $B^0 \to K^+\pi^-\gamma$
signal and 42\% of the $B^+ \to \Kpipi\gamma$ signal.

The $B$ decay signal is separated from background, first by
applying a requirement on $\DE$
and then by fitting the $\Mbc$ spectrum.
If we find multiple candidates
with $|\DE| < 0.5 \GeV$ and $\Mbc > 5.2 \GeV/c^2$
in the same event, we take the candidate
which gives the 
highest confidence
level when we fit the $K_X$ decay vertex
(best candidate selection).
We then select candidates with $-100\MeV < \DE < 75\MeV$,
which removes 19\% and 3\% of signal on the lower and
higher sides, respectively.
We define a $\DE$ sideband to be $100\MeV < \DE < 500\MeV$
at $\Mbc > 5.2 \GeV/c^2$,
in which we expect negligible signal contribution.

\begin{figure}
 \includegraphics[scale=0.867]{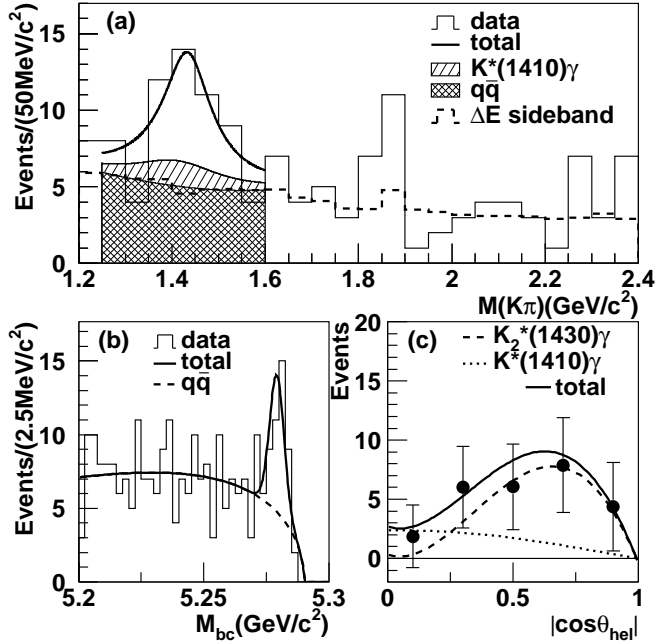}
 \caption{\label{fig:finalplot-k2st}
 (a) $M_{K\pi}$ (b) $\Mbc$ and (c) $|\coshel|$
 distributions for $B^0 \to \Kpi\gamma$ candidates.
 The unbinned ML fit results are shown in (a) and (c).
 The $\qq$ backgrounds are subtracted in (c).
 $\Mbc > 5.27 \GeV/c^2$ is applied in (a) and (c),
 and $1.25 \GeV/c^2 < M_{K\pi} < 1.6 \GeV/c^2$ is applied in (b)
 and (c).
 In (a), $\DE$ sideband data is scaled to the unbinned ML
 fit result and overlaid.
 }
\end{figure}

In the $B^0 \to \Kpi\gamma$ analysis,
we obtain the $M_{K\pi}$ distribution shown in
Fig.~\ref{fig:finalplot-k2st}(a).
We observe an excess
around $M_{K\pi}=1.4 \GeV/c^2$~\cite{footnote:d0pi0}.
The $\Mbc$ distribution with
$1.25 \GeV/c^2 < M_{K\pi} < 1.6 \GeV/c^2$ is shown
in Fig.~\ref{fig:finalplot-k2st}(b).
We fit
the $\Mbc$ distribution to extract the signal yield.
The distribution for the $\qq$ background
is modeled by an ARGUS function~\cite{ARGUS-function}
in which the shape is determined from
the $\DE$ data sideband.
The distribution for the signal component is modeled by
a Gaussian determined from signal MC calibrated
by $B^- \to D^0\pi^-$ data.
The signal yield is found to be $\YkpigammaOneFourNstr$
with a statistical significance of $\SNFkpigammaOneFourN\sigma$.
Here, the significance is defined as
$\sqrt{- 2 \ln ( {\cal L}(0) / {\cal L}_{\mathrm{max}} )}$, where
${\cal L}_{\mathrm{max}}$ is the maximum of the likelihood
and ${\cal L}(0)$ is the likelihood
for zero signal yield.

The observed signal may be explained as a mixture of
three components:
$B^0\to K_2^*(1430)^0\gamma$, $B^0\to K^*(1410)^0\gamma$
and non-resonant (N.R.) $B^0 \to K^+\pi^-\gamma$.
In order to separate these components,
we apply an unbinned maximum likelihood (ML) fit to $\Mbc$,
the cosine of the decay helicity angle
($\coshel$) and $M_{K\pi}$.
The expected $\coshel$ distributions are
$\sin^2 2\theta_{\rm hel}$, $\sin^2 \theta_{\rm hel}$
and uniform for these three components, respectively.
The PDFs for $\coshel$ and $M_{K\pi}$ are determined
from the $\DE$ sideband data for $\qq$ background,
from the corresponding MC samples for resonant components,
and from an inclusive $\bsgamma$ MC sample~\cite{Belle:2001:bsgamma}
for the non-resonant component.
The $\coshel$ PDFs for signals
are distorted up to 20\%
due to a non-uniform efficiency.
The validity of the method
is tested with
$B^- \to D^0\pi^-$ data and MC.

The fit results for $M_{K\pi}$ and $\coshel$ are
overlaid in Figs.~\ref{fig:finalplot-k2st}(a)
and \ref{fig:finalplot-k2st}(c),
and summarized in Table~\ref{tab:result}.
We find evidence for radiative decays via an intermediate tensor state,
$B^0 \to \Ktwost^0\gamma$.
The $K^*(1410)^0\gamma$ and non-resonant components are not
significant, so we set upper limits.
The 90\% confidence level upper limit $N$ is calculated from the
relation
$\int^{N}_{0} \mathcal{L}(n)dn = 0.9 \int^{\infty}_{0} \mathcal{L}(n)dn$,
where $\mathcal{L}(n)$ is the maximum likelihood with the signal yield
fixed at $n$.

We estimate the systematic error due to the fitting procedure as
follows.
For the signal shapes in the $\Mbc$ and $M_{K\pi}$ distributions,
we vary the mean and width parameters in the fit
within their experimental errors.
We also test the validity of
the background PDFs by replacing them with those obtained from a
$\qq$ MC sample. We assign the largest deviation in these tests as the
systematic error of the signal yield.

The event selection efficiency for $B^0 \to \Ktwost^0\gamma$ is
$\EFFktwostgammaN$ including the sub-decay branching fractions.
The error includes contributions from photon detection (2.8\%),
tracking (2.3\% per track), kaon identification (0.6\%),
pion identification (0.5\%),
event selection including likelihood ratio, $\pi^0/\eta$ veto
and best candidate selection (2.0\%)
and uncertainty of the sub-decay branching fractions (2.4\%).
Assuming an equal production rate for $B^0\bar{B}^0$
and $B^+B^-$,
this leads to a branching fraction of
$B^0 \to \Ktwost^0 \gamma$ of
$\BFktwostgammaNstr$.


The result agrees with the predictions based on a
relativistic form factor calculation
~\cite{kxgamma-prediction}.
Our result is also consistent with the CLEO measurement~\cite{CLEO:2000:radb}
when we neglect the non-resonant component and assume as they
did that the $K^*(1410)\gamma$ component is negligible.


In the $B^+ \to \Kpipi\gamma$ analysis,
we find additional background sources from a MC study.
Cross feed from $B \to K^*\gamma$
to $B^+ \to \Kpipi\gamma$ becomes
negligible after removing positively identified $B \to K\pi\gamma$
events.
The size of the cross feed from other $\bsgamma$ decays,
especially from those with a $\pi^0$ in the final state, is
estimated by using the inclusive $\bsgamma$ MC sample.
The contribution from the $b \to c$ background is
estimated by using a corresponding MC sample.

\begin{figure}
 \includegraphics[scale=0.867]{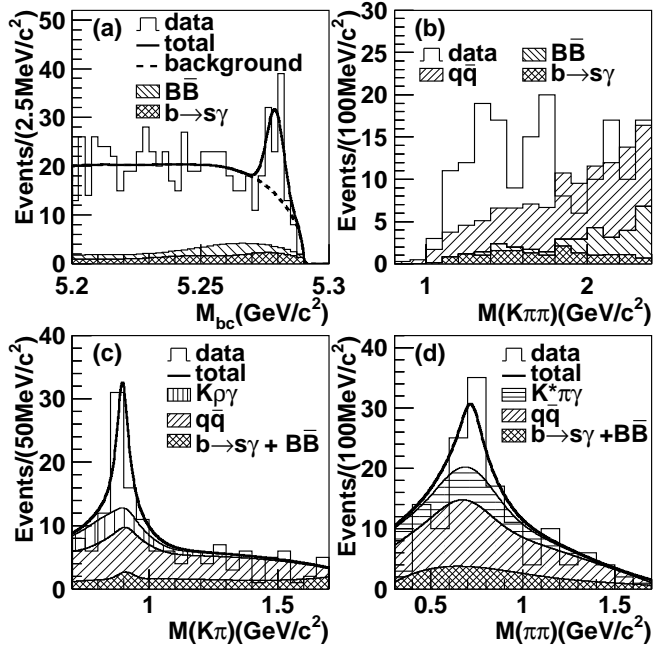}
 \caption{\label{fig:finalplot-kpipi}
 (a) $\Mbc$, (b) $\MKX$, (c) $M_{K\pi}$ and (d) $M_{\pi\pi}$
 distributions. The fit result of the $\Mbc$ distribution
 is shown in (a), while the result of the unbinned ML fit
 is shown in (c) and (d).
 $\Mbc > 5.27 \GeV/c^2$ is applied in (b), (c) and (d).
 }
\end{figure}

To extract the signal yield,
we fit the $\Mbc$ distribution shown in Fig.~\ref{fig:finalplot-kpipi}(a).
In addition to a Gaussian and an ARGUS function to describe
the signal and $\qq$ background components
obtained
using the same method as in the
$B \to K\pi\gamma$ analysis,
smoothed MC histograms for the $\bsgamma$ cross feed
and other $B$ meson decays
are used to model the $\Mbc$ shape,
where the normalizations are fixed assuming
the luminosity and the measured $\bsgamma$ branching
fraction~\cite{bf-bsgamma,Belle:2001:bsgamma}.
We find the signal yield of $\Ykpipigammastr$
with a $\SNFkpipigamma\sigma$ statistical significance.

The $\MKX$ distribution is shown in Fig.~\ref{fig:finalplot-kpipi}(b),
where the distribution for $\qq$ is obtained from
the $\Delta E$ sideband and is normalized
using the fit result.
We observe no signal excess above $1.8 \GeV/c^2$.
The $B^+ \to \Kpipi\gamma$ signal may be explained as a sum
of decays through
kaonic resonances such as
$B^+ \to K_1(1400)^+\gamma$ and $B^+ \to K^*(1680)^+ \gamma$.
The current statistics and the existence of a large number
of resonances prevent us from decomposing the
resonant substructure. However, it is still
possible to measure the $K^*\pi\gamma$ and $K\rho\gamma$ components
separately, as most of the resonances have
sizable decay rates through the $K^*\pi$ and $K\rho$ channels.

To find the composition of the signal,
we perform an unbinned ML fit
to $\Mbc$, $M_{K\pi}$ and $M_{\pi\pi}$
with three signal components ($K^*\pi\gamma$, $K\rho\gamma$
and non-resonant $K\pi\pi\gamma$) and
a $\qq$ background component.
In addition, the components from $\bsgamma$ cross feed and
from other $B$ meson decays are included in the fit
with fixed normalizations.
The $M_{K\pi}$ and $M_{\pi\pi}$ shapes for the $\qq$
background are determined from the $\DE$ sideband data,
and those for the other components are determined
from the corresponding MC samples.

In order to model the signal PDF
for the $K^*\pi\gamma$ component,
we use a mixture of
$B^+ \to K_1(1400)^+\gamma \to K^{*0}\pi^+\gamma$
and $B^+ \to K^*(1680)^+\gamma \to K^{*0}\pi^+\gamma$ MC.
The $K_1(1400)\gamma$ fraction of the mixture is determined to be
$\kstpiMCFRACkoneB$
by examining a background-subtracted $M_{K\pi\pi}$ distribution
for candidates with $|M_{K\pi} - M_{K^*}| < 75 \MeV/c^2$ ($K^*$ mass cut).
Likewise for the $K\rho\gamma$ PDF,
a mixture of
$B^+ \to K_1(1270)^+\gamma \to K^+\rho^0\gamma$
and $B^+ \to K^*(1680)^+\gamma \to K^+\rho^0\gamma$ MC
is used,
where the $K_1(1270)\gamma$ fraction is determined to be $\krhoMCFRACkoneA$
according to
a background-subtracted $M_{K\pi\pi}$ distribution for
candidates with $|M_{\pi\pi} - M_{\rho}| < 250 \MeV/c^2$
and $|M_{K\pi} - M_{K^*}| > 125 \MeV/c^2$ ($\rho$ mass cut).

\begin{table*}[t]
 \caption{\label{tab:result}%
 Measured signal yields, statistical significances, reconstruction
 efficiencies, branching fractions ($\Br)$ and
 90\% confidence level upper limits (UL) including systematic errors.
 The first and second errors are statistical and systematic, respectively.
 Efficiencies include the sub-decay branching
 fractions~\cite{PDG2000}.
 Efficiencies for $\Kpi\gamma$ and  $\Kpipi\gamma$ are based
 on a mixture of the measured sub-components.
 }
 \begin{ruledtabular}
  \catcode`;=\active \def;{\phantom{}}
  \begin{tabular}{ccccccc}
   Mode & Signal Yield & UL(yield) & Significance & Efficiency(\%)
   & $\Br$ $(\times 10^{-5})$ & UL $(\times 10^{-5})$ \\
   \hline
   $\Kpi\gamma$~\dag  & $\YkpigammaOneFourN$  & ---
   & $\SNFkpigammaOneFourN$\rlap{~\S}
   & $\EFFkpitotalNtbl$; & $\BFkpitotalNtbl$
   & --- \\
   $\Ktwost^0\gamma$ & $\YktwostgammaN$
   & --- & $\SNFktwostgammaN$
   & ;$\EFFktwostgammaNtbl$ & $\BFktwostgammaNtbl$ & --- \\
   $K^*(1410)^0\gamma$ & $\YkstAgammaNnosyst$
   & $\YULkstAgammaN$ & ---
   & ;$\EFFkstAgammaNtbl$ & --- & $\BFULkstAgammaNtbl$; \\
   $K^+\pi^-\gamma$ (N.R.)~\dag & $\YkpinresNnosyst$ &
   $\YULkpinresN$
   & --- & $\EFFkpinresNtbl$; & --- & ;$\BFULkpinresNtbl$ \\
   \hline
   $\Kpipi\gamma$~\ddag & $\Ykpipigamma$ & ---
   & $\SNFkpipigamma$\rlap{~\S} 
   & ;$\EFFkpipigammatbl$ & $\BFkpipigammatbl$ & --- \\
   $K^{*0}\pi^+\gamma$~\ddag & $\Ykstpigamma$ & ---
   & $\SNFkstpigamma$
   & ;$\EFFkstpigammatbl$ & $\BFkstpigammatbl$ & --- \\
   $K^+\rho^0\gamma$~\ddag & $\Ykrhogamma$ & $\YULkrhogamma$
   & $\SNFkrhogamma$
   & ;$\EFFkrhogammatbl$ & $\BFkrhogammatbl$ & ;$\BFULkrhogammatbl$; \\
   $\Kpipi\gamma$ (N.R.)~\ddag & $\Ynresnosyst$ & $\YULnres$ & ---
   & ;$\EFFnrestbl$ & --- & ;$\BFULnrestbl$ \\
   \hline
   $K_1(1270)^+\gamma$ & $\YkoneAgamma$
   & $\YULkoneAgamma$ & ---
   & ;$\EFFkoneAgammatbl$ & --- & ;$\BFULkoneAgammatbl$; \\
   $K_1(1400)^+\gamma$ & $\YkxOneFourgamma$ & $\YULkxOneFourgamma$
   & ---
   & ;$\EFFXkoneBgammatbl$ & --- & ;$\BFULXkoneBgammatbl$; \\
  \end{tabular}
 \end{ruledtabular}
 \leftline{\dag~$1.25 \GeV/c^2 < M_{K\pi} < 1.6 \GeV/c^2$
 \hspace{10mm}\ddag~$M_{K\pi\pi} < 2.4 \GeV/c^2$
 \hspace{10mm}\S~$\Mbc$ fit result
 }
\end{table*}

Figures~\ref{fig:finalplot-kpipi}(c) and \ref{fig:finalplot-kpipi}(d) show
the distributions and fit results for $M_{K\pi}$ and $M_{\pi\pi}$.
The selection efficiency is estimated from a MC sample
with the mixture of resonances used for the PDF determination.
We also consider other well-established resonances
~\cite{footnote:resonances}
which give slightly different efficiencies,
and assign the
difference in the result as a systematic error.
The signal yields, efficiencies
and the branching fractions are listed in
Table~\ref{tab:result}.
The total $B^+ \to \Kpipi\gamma$ branching fraction
is dominated by $B^+ \to K^{*0}\pi^+\gamma$ and $B^+ \to K^+\rho^0\gamma$;
the statistical significance for the sum of the two is
calculated to be $\SNFkstpiANDkrhogamma\sigma$
and the non-resonant component is consistent with zero.
We find evidence for the decay $B^+ \to K^{*0}\pi^+\gamma$
with a $\SNFkstpigamma\sigma$ significance,
while the $B^+ \to K^+\rho^0\gamma$ channel alone
yields only $\SNFkrhogamma\sigma$.
Systematic errors are evaluated using the same
procedures as in the $B \to K\pi\gamma$ analysis.

We also search for resonant decays by applying further
kinematical requirements.
We search for $B^+ \to K_1(1270)^+\gamma$ in the
$K^+\rho^0\gamma$ final state by applying the $\rho$ mass cut
and $|\MKX - M_{K_1(1270)}| < 100 \MeV/c^2$.
We find $\NSBOXkoneAgamma$ candidates
with a background
expectation of $\NBGkoneAgamma$ events.
To find $B^+ \to K_1(1400)^+\gamma$
in the $K^{*0}\pi^+\gamma$ final state,
we apply the $K^*$ mass cut and $|\MKX - M_{K_1(1400)}| < 200 \MeV/c^2$.
We obtain a sizable signal;
however we only provide upper limits due to a lack of ability
to distinguish these resonances.
The results are also listed in Table~\ref{tab:result}.


In conclusion, we have studied radiative $B$ decays
with the $\Kpi\gamma$ and $\Kpipi\gamma$ final states.
For $\Kpi\gamma$,
we consider $B^0 \to \Ktwost^0\gamma$,
$B^0 \to K^*(1410)^0\gamma$
and non-resonant components,
and find that only the first one is significant.
For $B^+ \to \Kpipi\gamma$, we observe the decay mode
and measure the branching fraction.
The branching fractions for $B \to K^*\pi\gamma$
and $K\rho\gamma$ are
consistent with the sum of predicted rates of
resonant decays~\cite{kxgamma-prediction}.
As listed in Table~\ref{tab:bflist},
we find $(35 \pm 8)\%$ of the total $B \to X_s\gamma$
decay is accounted for by the
$B \to K^*\gamma$, $B \to \Ktwost\gamma$,
and $B \to K\pi\pi\gamma$ final states.

\begin{table}[htbp]
 \caption{\label{tab:bflist}%
 Exclusive and inclusive branching fractions for the $\bsgamma$ process.
 Equal branching fractions are assumed for neutral and
 charged $B$ decays.
 Using isospin,
 the branching fraction of
 $B^+ \to K^{*+}\pi^0\gamma$ ($K^0\rho^+\gamma$) is
 assumed to be half (twice) of that for
 $B^+ \to K^{*0}\pi^+\gamma$ ($K^+\rho^0\gamma$).
 }
 \begin{ruledtabular}
  \catcode`;=\active \def;{\phantom{0}}
  \begin{tabular}{ccl}
   Mode & $\Br$ $(\times 10^{-5})$ & \multicolumn{1}{l}{Ref.} \\ \hline
   $B \to K^*\gamma$ & $;4.2 \pm 0.4$
   & \cite{CLEO:2000:radb,bf-K*gamma} \\
   $B \to \Ktwost\gamma$ (excluding $K^*\pi\gamma, K\rho\gamma$)
   & $;0.9 \pm 0.3$ & \\
   $B \to K^*\pi\gamma$  & $;3.1 \pm 1.0$ & \\
   $B \to K\rho\gamma$  & $;3.0 \pm 1.6$ & \\
   Sum of exclusive modes  & $11.2 \pm 2.1$ & \\
   $B \to X_s \gamma$ (inclusive) & $32.2 \pm 4.0$ &
   \cite{Belle:2001:bsgamma,bf-bsgamma} \\
  \end{tabular}
 \end{ruledtabular}
\end{table}


We wish to thank the KEKB accelerator group for the excellent
operation of the KEKB accelerator.
We acknowledge support from the Ministry of Education,
Culture, Sports, Science, and Technology of Japan
and the Japan Society for the Promotion of Science;
the Australian Research Council
and the Australian Department of Industry, Science and Resources;
the National Science Foundation of China under contract No.~10175071;
the Department of Science and Technology of India;
the BK21 program of the Ministry of Education of Korea
and the CHEP SRC program of the Korea Science and Engineering Foundation;
the Polish State Committee for Scientific Research
under contract No.~2P03B 17017;
the Ministry of Science and Technology of the Russian Federation;
the Ministry of Education, Science and Sport of the Republic of Slovenia;
the National Science Council and the Ministry of Education of Taiwan;
and the U.S.\ Department of Energy.




\begin{thebibliography}{99}
\bibitem{footnote:charge-conjugation}
	 The charge conjugated modes are implicitly included.
\bibitem{footnote:d0pi0}
	 We expect $3 \pm 1$ $\bar{B}^0 \to D^0\pi^0$ background
	 which may account for the excess around $M_{K\pi}=1.85 \GeV/c^2$.
\end{thebibliography}

%
%

\end{document}